\documentclass[10pt,a4paper]{article}
\usepackage[position=b,font=normalsize]{subcaption}
\usepackage[utf8]{inputenc}
\usepackage{epstopdf}
\usepackage{layout}
\usepackage{placeins}
\usepackage{amsmath}
\usepackage{geometry}
\usepackage{graphicx}
\usepackage{upgreek}
\usepackage{authblk}
\usepackage{hyperref}

\newcommand{\JT}{\mathcal{A}}
\newcommand{\JS}{\mathcal{A}}
\newcommand{\s}{\mathrm{s}}
\renewcommand{\i}{\mathrm{r}}
\newcommand{\p}{\mathrm{p}}
\newcommand{\q}{\mathrm{q}}
\renewcommand{\c}{\mathrm{c}}

\newcommand{\A}{\hat{a}}
\newcommand{\Ad}{\hat{a}^{\dagger}}

\newcommand{\HC}{\,\mathrm{H.C.}}
\newcommand{\ud}{\,\mathrm{d}}
\newcommand{\mean}[1]{ \left \langle #1 \right \rangle}
\newcommand{\e}{\mathrm{e}}
\newcommand{\rmi}{\mathrm{i}}
\newcommand{\cc}{\mathrm{c.c.}}
\newcommand{\Tr}{\mathrm{Tr}}
\newcommand{\M}{\hat{M}}
\newcommand{\ket}[1]{|#1\rangle}
\newcommand{\LPo}{$\mathrm{LP}_{01}$}
\newcommand{\LPt}{$\mathrm{LP}_{11}$}
\newcommand{\eref}[1]{(\ref{#1})}
\begin{document}

\title{\textbf{Spectrally pure heralded single photons by spontaneous four-wave mixing in a fiber: \\ reducing impact of dispersion fluctuations}}
\author[1]{Jacob G. Koefoed}
\author[1]{Søren M. M. Friis}
\author[1]{Jesper B. Christensen}
\author[1]{Karsten~Rottwitt}
\affil{\small Department of Photonics Engineering, Technical University of Denmark}

\date{}
\maketitle



\pagestyle{plain}
\begin{abstract}
We model the spectral quantum-mechanical purity of heralded single photons from a photon-pair source based on nondegenerate spontaneous four-wave mixing taking the impact of distributed dispersion fluctuations into account. The considered photon-pair-generation scheme utilizes pump-pulse walk-off to produce pure heralded photons and phase matching is achieved through the dispersion properties of distinct spatial modes in a few-mode silica step-index fiber. We show that fiber-core-radius fluctuations in general severely impact the single-photon purity. Furthermore, by optimizing the fiber design we show that generation of single photons with very high spectral purity is feasible even in the presence of large core-radius fluctuations. At the same time, contamination from spontaneous Raman scattering is greatly mitigated by separating the single-photon frequency by more than 32 THz from the pump frequency.
\end{abstract}

\section{Introduction}
Many applications of quantum photonics, such as quantum cryptography\cite{gisin2007quantum, ekert1991quantum} and quantum computation\cite{knill2001scheme, ladd2010quantum}, rely critically on the ability to design single-photon sources with high generation rate, low noise and photon indistinguishability. An important subset of single-photon sources are pair sources based on the nonlinear optical effects of spontaneous parametric down-conversion (SPDC)\cite{tanzilli2001highly, banaszek2001generation, u2004efficient} or spontaneous four-wave mixing (SpFWM), the latter of which has been demonstrated in dispersion-shifted fibers\cite{li2005optical}, photonic crystal fibers (PCFs)\cite{sharping2004quantum, rarity2005photonic, cohen2009tailored} and silicon waveguides\cite{sharping2006generation, xiong2011slow}. Although such sources are indeterministic in nature, one of the photons in the pair can be used to herald the presence of the other\cite{fasel2004high,mcmillan2009narrowband}. However, this heralding process generally projects the heralded photon into an impure, classical mixture of frequency states, resulting in distinguishable photons. \par 
Many important applications, such as linear optical quantum computing, rely on two-photon interference\cite{hong1987measurement}, which is limited by single-photon indistinguishability\cite{walmsley2005toward, mosley2008heralded}. For this reason, much attention has been given to the challenge of generating spectrally uncorrelated photon pairs without the use of narrow spectral filtering in e.g. photonic crystal fibers (PCFs)~\cite{cohen2009tailored, francis2016all}, resulting in heralded quantum-mechanically pure single photons at large frequency separation from the pump. Schemes not relying on spectral filtering to achieve pure photons have the advantage of lower losses, and sufficient spectral separation from the pump results in low contamination from spontaneous Raman scattering (SpRS). In most glass-based platforms, Raman scattering is a significant source of noise photons~\cite{dyer2009high} and it affects the single-photon purity~\cite{koefoed2017effects}. While these prospects are attractive, experiments in PCFs have resulted in purities below theoretical estimates\cite{halder2009nonclassical, clark2011intrinsically}, and it has been shown that a significant reduction in purity can result from parasitic nonlinear phase modulation~\cite{bell2015effects}. Since high pre-filtering purity of the heralded photon depends critically on waveguide dispersion properties, inhomogeneity along the waveguide length is another source of purity reduction, which has been experimentally verified in the case of PCFs~\cite{cui2012spectral, francis2016characterisation}. However, little effort has been devoted to making photon-pair sources less vulnerable to dispersion fluctuations and understanding the limit that it imposes on e.g. fiber length and pump pulse durations.\par
In this work, we develop a general model that describes the joint state of photon pairs under the effects of stochastic variations of waveguide parameters. This model is used to characterize the impact of random fiber core-radius fluctuations (CRF) in the case where a simple silica step-index fiber (SIF) is used for pure single-photon generation. The scheme we consider for pure photon generation was recently proposed\cite{christensen2016temporally} and relies on two pumps travelling at different group velocities due to modal nondegeneracy. The gradual variation of the nonlinear interaction strength enables the generation of highly factorable two-photon states. The scheme was originally proposed for polarization modes, but here we demonstrate that a multi-mode fiber platform is equally feasible. Although schemes relying on specific dispersion properties of the fiber are often realized in PCFs, we show that a simple SIF carrying higher-order modes can exhibit sufficient flexibility to generate pure single photons. We optimize the parameters of a silica SIF to enable an intermodal four-wave-mixing process that is very robust to fluctuations in fiber geometry. Additionally, spontaneous Raman scattering is minimized at the single-photon frequencies due to a large frequency separation from the pumps, facilitated by the different propagation constants of higher-order fiber modes. Lastly, we discuss limitations on fiber length and pump pulse durations as well as the impact of the correlation length of CRF in the fiber. Although our proposed implementation of stochastic waveguide geometry variations in the description of the joint state of photon pairs is applied to a silica fiber, a similar analysis can be applied to any platform used for pure single-photon generation through SpFWM.

\section{Theory}

\subsection{Joint state of photon pairs with fluctuating fiber parameters}
We consider a photon-pair generation process through SpFWM whereby two pulsed classical pump fields, $A_\p$ and $A_\q$, propagate through a waveguide. We assume all fields are linearly polarized along the unit vector $\vec{e}$ and decompose the classical pump fields as
\begin{equation}
\label{pump_fields}
\vec{E}_j = \frac{1}{2}\vec{e} F_j(x,y,z)  \sqrt{\frac{2}{n_j\epsilon_0 c }} \left [ A_j(z,t)\e^{ -\rmi\omega_{0 j}t} +\cc \right ], \quad j = \p,\q,
\end{equation}
where $F_j(x,y,z)$ is the mode profile, normalized such that $|F_j(x,y)|^2$ integrates to unity over the waveguide cross section, and allowed a slow dependence on the longitudinal waveguide coordinate. Additionally, $n_j$ is the refractive index at the central angular frequency $\omega_{0j}$, and $A_j(z,t)$ is the slowly-varying envelope with $|A_j|^2$ representing optical power. In the case where the fields are spectrally narrow compared to their central frequencies, we may, to a very good approximation, quantize the signal and idler vector operator fields as
\begin{equation}
\vec{\hat{E}}_j(z,t) = \frac{1}{2}\vec{e}F_j(x,y,z)\e^{-\rmi\omega_{j 0}t} \sqrt{ \frac{2\hbar \omega_{j 0}}{n_{j 0}\epsilon_0  c }} \A_j(z,t) + \HC,\quad j =\s,\i,
\end{equation}
where $\A_j(z,t)$ is the slowly-varying field operator for field $j = \s,\i$, which is the (inverse) Fourier transform of the usual annihilation operator $\A_j(z,\omega)$, and with equal-position commutator \\ $[\A_j(z,t), \Ad_i(z,t')] = \delta_{ij}\delta(t-t')$. We model the evolution of the photon-pair state in the interaction picture through the interaction momentum operator~\cite{abram1991quantum,sinclair2016effect,koefoed2017effects}
\begin{equation}
\label{H_start}
\M_\mathrm{int}(z) = 2\hbar \gamma(z) \int_{-\infty}^\infty \ud t A_\p(z,t) A_\q(z,t) \Ad_\s(z,t) \Ad_\i(z,t) + \HC,
\end{equation}
governing spatial evolution of the fields such that $-i\hbar \partial_z \ket{\psi} = \M_\mathrm{int}(z) \ket{\psi}$. The factor of 2 results from the nondegenerate pumps and we have allowed the nonlinear interaction strength
\begin{equation}
\gamma(z) = \frac{3\chi^{(3)} \sqrt{\omega_{\s 0} \omega_{\i 0}} f_{\p \s \p \i}(z)}{4 \epsilon_0 c^2  \sqrt{n_\p n_\q n_\s n_\i}},
\end{equation}
to depend on $z$ through the nonlinear mode overlap
\begin{equation}
f_{\p\s\q\i}(z) = \iint \ud x \ud y F_\p(x,y,z) F_\s^*(x,y,z) F_\q(x,y,z) F_\i^*(x,y,z).
\end{equation}
Even though nonlinear phase-modulation can be included in this model, its effect has been analyzed previously~\cite{christensen2016temporally} and since we expect no relevant interaction with the effects considered in this work we neglect it along with intrapulse group-velocity dispersion (GVD). Under the assumption that the longitudinal variations of parameters happen on a much longer length scale than the wavelength of light, so that we may neglect $z$-derivatives compared to factors of $\beta_1$, all fields evolve according to a simple nonlinear Schrödinger equation
\begin{equation}
\partial_z \mathcal{F}_j(z,t) + \beta_{1j}(z)\partial_t \mathcal{F}_j(z,t) = i \beta_{0j}(z) \mathcal{F}_j(z,t), \quad \mathcal{F}_j = A_\p, A_\q,\A_\s, \A_\i.
\end{equation}
Here $\beta_{0j}(z)$ and $\beta_{1j}(z)$ are the position-dependent propagation constant and inverse group velocity of field $j$, respectively. The solution to this equation takes the form
\begin{equation}
\label{field_solution}
\mathcal{F}_j(z,t) =\mathcal{F}_j\left (0,t - \int_0^z \ud z' \beta_{1j}(z')\right ) \exp\left (i \int_0^z \ud z'\beta_{0j}(z')\right ), \quad \mathcal{F}_j = A_\p, A_\q,\A_\s, \A_\i,
\end{equation}
which is easily verified by insertion. The joint temporal amplitude (JTA) $\JT(t_\s,t_\i)$ presents a way to conveniently express the unnormalized two-photon part of the quantum state
\begin{equation}
\label{eq:state}
\ket{\psi} = \iint \ud t_\s \ud t_\i \JT(t_\s,t_\i) \ket{t_\s} \ket{t_\i},
\end{equation}
expressed in the time basis $\ket{t_j} = \Ad_j(L,t_j) \ket{\mathrm{vac}}$, where $ \ket{\mathrm{vac}}$ is the vacuum state. Thus, the JTA gives the distribution of temporal modes and holds information about the temporal correlations between signal and idler photons. The joint spectral amplitude (JSA) $\JS(\omega_\s,\omega_\i)$ plays the same role for spectral modes and is simply the 2D Fourier transform of the JTA. Eq. \eref{eq:state} indicates that to first order in the pair-production rate the JTA is given by
\begin{equation}
\JT(t_\s,t_\i) = \mean{\A_\s(L,t_\s)\A_\i(L,t_\i)   \left (\frac{i}{\hbar}\int_0^L \ud z \M_\mathrm{int}(z) \right )},
\end{equation}
where $L$ is the waveguide length and expectation values are with respect to a vacuum input. Using Eqs. \eref{H_start} and \eref{field_solution}, this can be shown to be
\begin{eqnarray}
\label{joint_state_z}
\JT(t_\s,t_\i) &=& 2i \gamma(z_\c) A_\p\left (0,t_\c - \int_0^{z_\c} \ud z' \beta_{1\p}(z')\right ) A_\q\left (0,t_\c - \int_0^{z_\c} \ud z' \beta_{1\q}(z')\right ) \notag\\
&&\times \exp\left (i \int_0^{z_\c} \ud z' \Delta\beta_0(z')\right ) \Theta(z_\c)\Theta(L - z_\c),
\end{eqnarray}
where $\Delta \beta_0(z) = \beta_{0\p}(z) + \beta_{0\q}(z) - \beta_{0\s}(z) - \beta_{0\i}(z)$ is the position-dependent phase mismatch. Here, $\Theta$ is the Heaviside function and the collision coordinates $(z_\c,t_\c)$ are given by the solutions to
\begin{subequations}
\begin{eqnarray}
t_\c = t_\s - \int_{z_\c}^L \ud z' \beta_{1\s}(z'), \\
t_\c = t_\i - \int_{z_\c}^L \ud z' \beta_{1\i}(z').
\end{eqnarray}
\end{subequations}
These equations describe the stochastic creation time $t_\c$ and position $z_\c$ of the photon pair inside the waveguide. \par
An important measure for single-photon sources is their achievable interference visibility $V = \Tr(\hat{\rho}_\s^{(1)}\hat{\rho}_\s^{(2)})$, from sources 1 and 2. For photon-pair sources the single-photon signal state heralded from the total state $\hat{\rho} = |\psi \rangle \langle \psi |$ is $\hat{\rho_\s} = \Tr_\i(\hat{\rho})$~\cite{mosley2008heralded}. In this case, the visibility can be expressed in terms of the individual JTAs as
\begin{equation}
\label{eq:visibility}
V = \frac{1}{R_1 R_2}\iint \ud t_\s \ud t_\s' \left (\int \ud t_\i \JT_1(t_\s, t_\i) \JT_1^*(t_\s',t_\i) \right ) \left (\int \ud t_\i' \JT_2(t_\s', t_\i') \JT_2^*(t_\s,t_\i') \right ),
\end{equation}
where $R_j = \iint \ud t_\s t_\i |\JT_j(t_\s, t_\i)|^2$, $j=1,2$ are the photon-pair generation probabilities. If the two sources are identical, or if two heralded photons from the same source are interfered, so that $\JT_1 = \JT_2$, the visibility reduces to the heralded purity $P$. In general $0 \leq V \leq P \leq 1$ with the last equality being valid if $\JT$ is factorable so that $\JT(t_\s,t_\i) = \JT_\s(t_\s)\JT_\i(t_\i)$, which can be seen from Eq. \eref{eq:visibility} by inserting two identical and factorable JTAs. Alternatively, the heralded purity can also be found through a Schmidt decomposition~\cite{grice2001eliminating} of the JTA of the form $\JT(t_\s,t_\i) = \sum_n \lambda_n f_n(t_\s) g_n(t_\i)$, from which the purity is calculated as~\cite{u2006generation}
\begin{equation} \label{eq_Purity}
P = \frac{\sum_n |\lambda_n|^4}{\left( \sum_n |\lambda_n|^2\right )^2},
\end{equation}
where the denominator is the generation probability, which is usually kept below 0.1 to avoid multipair emissions.

\subsection{Factorability of joint state without fluctuations}
\label{sec:factorability}
In the absence of fluctuations in the waveguide geometry, the joint state Eq. \eref{joint_state_z} becomes
\begin{equation}
\label{joint_state_nofluc}
\JT(t_\s,t_\i) = 2i \gamma A_\p(0,t_\c -  \beta_{1\p}z_\c ) A_\q(0,t_\c - \beta_{1\q}z_\c ) \Theta(z_\c)\Theta(L - z_\c),
\end{equation}
and explicit expression can be found for the collision coordinates:
\begin{equation}
z_\c = L - \frac{t_\s - t_\i}{\beta_{1\s}-\beta_{1\i}}, \qquad\qquad  t_\c = \frac{\beta_{1\s} t_\i - \beta_{1\i} t_\s}{\beta_{1\s} - \beta_{1\i}}.
\end{equation}
The Heaviside functions in Eq.~\eref{joint_state_nofluc} are nonfactorable functions in $(t_s, t_r)$, and therefore introduce unwanted temporal correlation in the JTA. Physically, this effect occurs due to the abrupt change in nonlinear interaction at the waveguide endpoints. While this is inherent to single-pump configurations where the interaction is uniform, the effect can be completely avoided using a configuration of nondegenerate pulsed pumps walking through each other inside the waveguide.   
To further investigate the JTA in Eq.~\eref{joint_state_nofluc} we here, and for the remainder of this work, assume Gaussian pumps of the form 
\begin{equation}
A_j(0,t) = \sqrt{P_j} \exp\left (-\frac{(t-\Delta t_j)^2}{2 T_j^2}\right ) , \quad j = \p,\q,
\end{equation}
where $P_j$ is the pulse peak power, $T_j$ is the pulse duration, and $\Delta t_j$ is an initial temporal displacement. We furthermore assume a full pump collision, meaning that the slower pump enters the fiber before the faster pump and that the pumps pass completely through each other temporally, so that $|\beta_{1\p} - \beta_{1\q}|L \gg |\Delta t_\p - \Delta t_\q| \gg \sqrt{T_\p^2 + T_\q^2}$. In this case, the joint state is factorable if\cite{fang2013state, christensen2016temporally}
\begin{equation}
\label{purity_criter}
T_\q^2(\beta_{1\p} - \beta_{1\i})(\beta_{1\p} - \beta_{1\s}) 
+ T_\p^2 (\beta_{1\q} - \beta_{1\i})(\beta_{1\q} - \beta_{1\s}) = 0,
\end{equation}
which depends only on the relative inverse group velocities (IGV) between the four fields and the pump pulse durations. A particularly simple solution to Eq.~\eref{purity_criter} occurs when the pumps and sidebands pairwise copropagate such that, for example, $\beta_{1\p} = \beta_{1\i} \neq \beta_{1\q} = \beta_{1\s}$. Typically, this condition is approximately met in situations where the two pumps are excited in two different spatial waveguide modes. In such an intermodal four-wave-mixing process, the phase-matching condition takes the form of a parallelogram in $(\omega$-$\beta_{1})$-space as described in Sec. \ref{sec:parallelogram}, and thereby directly leads to the desired pairwise group-velocity matching of Eq. \eref{purity_criter}. \par
In this work, we restrict our analysis to the case where the two initial pump envelopes are temporally identical, except for their initial temporal separation. We show that very high purities are still achievable, even with this restriction. To ensure complete walk-through it is sufficient to choose an initial pump separation of $|\Delta t_\p - \Delta t_\q| = 4 \sqrt{T_\p^2 + T_\q^2} = 4\sqrt{2} T_p$ and a fiber length $L = 2|\Delta t_\p - \Delta t_\q|/|\beta_{1\p} - \beta_{1\q}|$, so that the pump pulses coincide temporally at the fiber midpoint.

\begin{figure}
\centering
\includegraphics[scale=1]{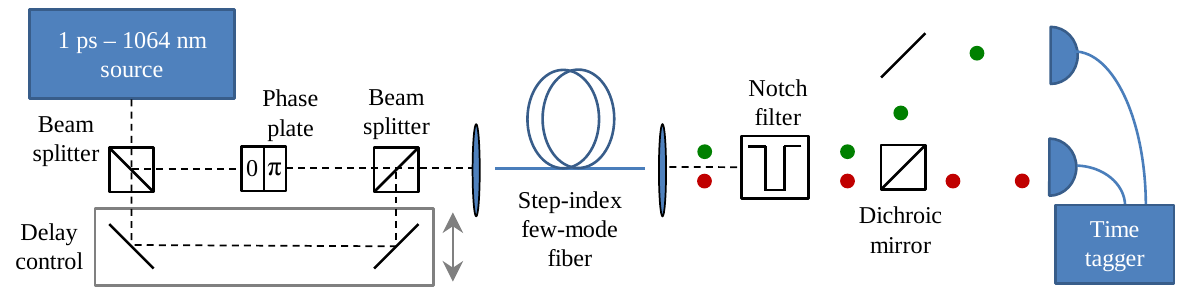}
\caption{Possible setup for pure single-photon generation; the green and red dots are single photons at different wavelengths that are generated through SpFWM in the few-mode SIF. }
\label{fig:setup}
\end{figure}

This scheme for generating pure heralded single photons is exemplified in Fig. \ref{fig:setup}. The radiation from a pulsed source (dashed line), for example a 1-ps 1064-nm laser as we use later, is divided in two; the first beam goes through a phase plate that changes the wave front such the LP$_{11}$-mode is excited when entering the fiber; the second goes through a delay stage such that the pulse that excites the LP$_{01}$-mode, which has a higher group velocity than the LP$_{11}$-mode, enters the fiber later. In the few-mode SIF, the LP$_{01}$ pump pulse pass through the LP$_{11}$ pulse and with a probability depending on the pulse peak power, one photon from each of the pump pulses decay to a signal and idler photon, marked with red and green dots, respectively, at the output. A notch filter removes the pump from the beam path and a dichroic mirror separates the signal and idler photons.

\section{Photon-pair generation in a two-mode silica fiber}

In this section, we exemplify a realization of the scheme for generating spectrally pure heralded single photons in a two-mode fiber, including an explanation of the phase matching of the LP$_{01}$ and \LPt-modes and a model of stochastic core radius variation. We show that a purity of unity is obtainable, and afterwards we take into account CRF and show that the high purity is destroyed even by small degrees of fluctuations. We consider a weakly guiding SIF in which the propagation constants of the guided modes are found by solving the scalar wave equation\cite{Okamoto2006}. These linearly polarized solutions are denoted with indexes $m$ and $l$, establishing the angular and radial dependence, respectively, such that all fiber modes are identified by the usual designation LP$_{ml}$.

\subsection{Two-mode phase matching} \label{sec:parallelogram}
\begin{figure}
\centering\begin{subfigure}{0.49\linewidth}  \centering
\includegraphics[scale=1]{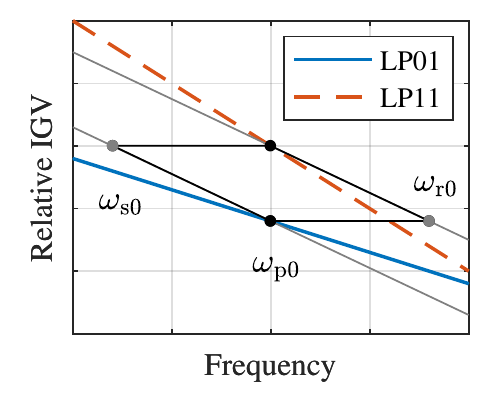}
\caption{}
\end{subfigure} 
\begin{subfigure}{0.49\linewidth}  \centering
\includegraphics[scale=1]{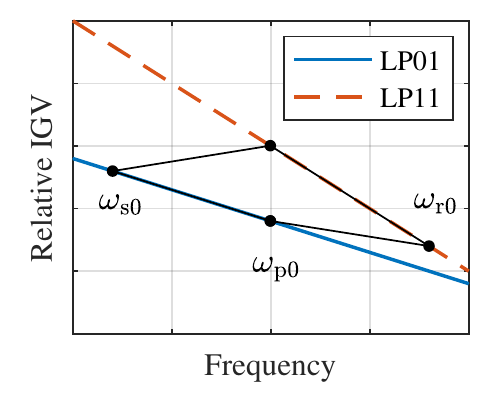}
\caption{}
\end{subfigure}
\caption{(a) Diagram of relative IGV of the \LPo-mode (solid blue) and \LPt-mode (dashed red); the black dots denote the IGV of the pump modes at the central pump frequency; the solid gray lines lie on the pump IGV at the pump frequency but their slope value is the average of the \LPo and \LPt slope values; the gray dots denote the points on the grey lines that form a parallelogram with the black pump dots; (b) the same relative IGV diagram as in (a) but showing the skewed parallelogram of phase matching that satisfies Eq. \eref{eq_DBapprox}.} \label{fig_RIGV_sketch}
\end{figure}

The phase-matching properties of two-mode four-wave mixing have been studied recently \cite{Essiambre2013,Friis2016,Parmigiani2016}, and in the special case of photon-pair generation in the LP$_{01}$ and \LPt-modes, the phase mismatch is
\begin{equation} \label{eq_DB}
\Delta \beta = \beta^{(01)}(\omega_{\p0}) + \beta^{(11)}(\omega_{\p0}) - \beta^{(01)}(\omega_{\s0}) -\beta^{(11)}(\omega_{\i0}),
\end{equation}
where $\beta^{(\mu)}(\omega_{j0})$ denotes the propagation constant in mode $\mu$ at angular frequency $\omega_{j0}$. In this configuration, a pump at angular frequency $\omega_{\p 0}$ distributed in the \LPo and \LPt-modes generates photons with central frequencies $\omega_{\s 0}$ and $\omega_{\i 0} > \omega_{\s 0}$ in the \LPt-mode and \LPo-mode, respectively. In terms of the pump-idler frequency separation $\Omega = \omega_{\i 0} - \omega_{\p 0}$, the phase mismatch can be approximated as
\begin{equation} \label{eq_DBapprox}
\Delta \beta \approx \left( \beta_1^{(01)} - \beta_1^{(11)} + \frac{\beta_2^{(01)} + \beta_2^{(11)}}{2} \Omega  \right) \Omega,
\end{equation}
by expanding each propagation constant in a Taylor series around $\omega_{\rm p0}$ to second order and exploiting the energy conservation $2\omega_{\p0} = \omega_{\s 0} + \omega_{\i 0}$, where $\beta_i^{(\mu)}$ is the $i$'th order dispersion coefficient in mode $\mu$ at $ \omega_{\p 0}$. In this simple case of negligible third- and higher-order dispersion terms, Eq. \eref{eq_DBapprox} has a simple graphical interpretation: If the inverse group velocities are plotted versus frequency and $\beta_2^{(\mu)}$ for $\mu \in \{ 01,11\}$ are replaced with their average values, the phase mismatch is zero at the $\omega_{\s 0}$-value where the IGVs of all four waves form a parallelogram, as illustrated in Fig. \ref{fig_RIGV_sketch}(a). Thus, if the GVDs of the two modes are equal, i.e. $\beta_2^{(01)} = \beta_2^{(11)}$, the IGV of the four waves inherently form a parallelogram and Eq. \eref{purity_criter} implies that phase matching automatically yields factorable joint states. 
Unfortunately, it turns out that it is difficult to achieve $\beta_2^{(01)}\approx \beta_2^{(11)}$ in a SIF, while also keeping the frequency separation of the IGV lines in Fig. \ref{fig_RIGV_sketch}(a) larger than the Raman response spectrum (the last peak is located at 32 THz) to avoid SpRS noise contamination at the signal and idler frequencies. Hence, the phase-matched IGVs of the four fields form a skewed parallelogram as shown in Fig. \ref{fig_RIGV_sketch}(b), which may reduce the purity of the heralded photons. However, as we show in Sec. \ref{sec:Joint_state_TMF}, this reduction is not necessarily significant. Phase matching can be achieved at very large frequency shifts outside the Raman response spectrum if higher-order dispersion terms are taken into account, and these solutions also have skewed parallelograms when plotting the IGV, as shown later in Fig. \ref{fig_cartoon}.


\subsection{Model for core-radius fluctuations}\label{sec:fluctuations}

To characterize the impact of geometric fluctuations on the scheme presented above, we define a stochastic model that describes these fluctuations. Studies on random variations in the dispersion properties of fibers indicate that fluctuations in index contrast have a small impact compared to fluctuations in core radius\cite{kuwaki1990evaluation}. We therefore neglect fluctuations in doping concentration and model the random variation of the core radius through the fiber with a Gaussian stochastic process\cite{karlsson1998four,yaman2004impact}, where the core radius is perturbed randomly away from the design target value $a_0$ with a correlation length $l_\mathrm{corr}$ so that
\begin{equation}
\label{langevin}
a(z) = a_0 + \int_{-\infty}^z \ud z' \hat{N}(z') \e^{-(z-z')/l_\mathrm{corr}},
\end{equation}
where $\hat{N}(z)$ is a Langevin noise source and the lower limit is taken to be $-\infty$ to account for the fact that the core radius at the beginning of the fiber, $z = 0$, is also randomly distributed. The dimensionless noise correlations are taken to be
\begin{equation}
\mean{\hat{N}(z)\hat{N}(z')} = \frac{2\sigma_a^2}{l_\mathrm{corr}} \delta(z-z'),
\end{equation}
so that the variance in the core radius is $\mathrm{Var}[a(z)] = \sigma_a^2$ at any point in the fiber. In real fibers, the magnitude of these fluctuations are up to 1 \%\cite{kuwaki1990evaluation,karlsson1998four}. The correlation length varies depending on the exact manufacturing process, but is typically on a scale of meters\cite{karlsson1998four,yaman2004impact}. As we show later, the most important fluctuations are those happening on a length scale comparable to the nonlinear interaction length. In the following, we use the realistic value $l_\mathrm{corr} = 1~\mathrm{m}$, and later investigate the exact dependence on this parameter. Numerically, we simulate the deviation $\Delta a(z) = a(z) - a_0$ as a sampling of the Langevin process of Eq. \eref{langevin}, i.e. $\Delta a_{n+1} = \alpha \Delta a_n + \delta a_n$ with $\delta a_n \sim \mathcal{N}(0,\sigma_a^2 (1-\alpha^2))$, $\alpha = \exp(-\Delta z/l_\mathrm{corr})$ and $\Delta a_0 \sim \mathcal{N}(0,\sigma_a^2)$, where $\mathcal{N}(\mu,\sigma^2)$ is the normal distribution with mean $\mu$ and variance $\sigma^2$.


\subsection{Joint state from a two-mode fiber}
\label{sec:Joint_state_TMF}

\begin{figure}[ht]
\centering
\includegraphics[scale=1.0]{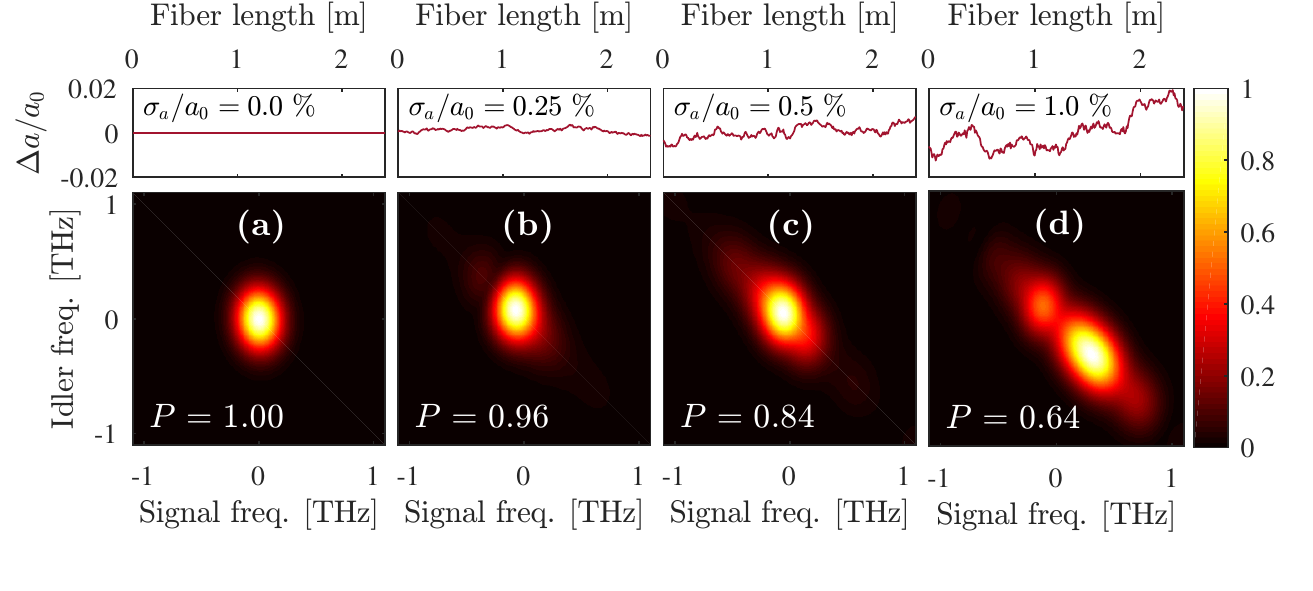}
\caption{The relative core radius variation along the fiber length (top) and the normalized absolute value of the resulting JSA (bottom) for \textbf{(a)} $\sigma_a = 0$, \textbf{(b)}  $\sigma_a/a_0 = 0.25~\%$, \textbf{(c)} $\sigma_a/a_0 = 0.5~\%$, and \textbf{(d)} $\sigma_a/a_0 = 1.0~\%$.}
\label{fig:JSA_fluctuations}
\end{figure}

As an example, we consider a SIF with core radius $a_0 = 4.0~\upmu\mathrm{m}$ and a Germanium-doped core with a doping concentration of 6.7 \%, which corresponds to a refractive index difference of $\Delta n  = 9.9\cdot 10^{-3}$ at 1064 nm between core and cladding \cite{Adams_1981}; the pump is fixed at $\lambda_{\rm p0} = 1064$~nm, where a large range of high repetition sources are readily available, and the phase-matched signal and idler wavelengths become $\lambda_{\rm s0} = 1216$~nm and $\lambda_{\rm r0} = 945.5$~nm, respectively. We use pump pulse widths of 1~ps and a fiber length of 2.4~m with an initial pump pulse temporal separation that ensures full walk-off as explained in the theory section. Equation \eref{langevin} is used to find the fiber parameters along the waveguide and Eq.~\eref{joint_state_z} is used to calculate the JSA. Fig.~\ref{fig:JSA_fluctuations}(a) shows the result in the absence of any CRF, i.e. with $\sigma_a = 0$: The state shows no correlation between the signal and idler frequencies, which is a signature of perfect factorability and hence a purity of 1.00, calculated by Eq. \eref{eq_Purity}. Figures \ref{fig:JSA_fluctuations}(b)--\ref{fig:JSA_fluctuations}(d) show the JSA for increasing degrees of CRF, and the top plots show the simulated relative core-radius variation through the fiber. The very pure state in Fig. \ref{fig:JSA_fluctuations}(a) is slightly distorted with an accompanying reduction in purity to 0.96 in Fig. \ref{fig:JSA_fluctuations}(b) by small fluctuations with $\sigma_a/a_0 = 0.25~\%$. At $\sigma_a/a_0 = 0.5~\%$ (the maximum variation of the core radius is well below 1~\%) a more significant distortion is observed in Fig. \ref{fig:JSA_fluctuations}(c) leading to a purity of $0.84$, which is below existing fiber sources~\cite{francis2016all}. Finally, in Fig. \ref{fig:JSA_fluctuations}(d) for $\sigma_a/a_0 = 1.0~\%$ the formerly uncorrelated joint state is completely destroyed resulting in significant spectral correlation and a low purity of $0.64$.
This shows that even though the scheme of Sec. \ref{sec:factorability} can be realized in a simple silica SIF by exciting the pump in two different spatial modes, the very high purity is destroyed by CRF smaller than $\sigma_a/a_0 = 0.5~\%$.  Additionally, the movement of the central peak of the JSA in Fig. \ref{fig:JSA_fluctuations} indicates that the distortion is mainly due to fluctuations in the phase-matched frequencies. We expect that random variations in IGV, which are expected to change the shape of the JSA, only have a small destructive effect while variation in the nonlinear coupling due to changes in fiber-mode distributions is almost completely negligible.

\FloatBarrier

\section{Reducing impact of core-radius fluctuations}\label{sec:results}

In light of the results of the previous section, we aim to design a SIF that reduces the impact of CRF. Recently, a more comprehensive design of a single-mode nonlinear fiber with such a property has been proposed \cite{Kuo2012} but multi-mode SIFs have other degrees of freedom that we may utilize to this end. In the following, we present a fiber design that effectively suppresses the impact of CRF and document its performance by calculating the heralded purity and Hong-Ou-Mandel visibility, which we show to be significantly improved from the fiber discussed above.

\subsection{Optimizing the fiber design}
\label{sec:design}
A silica SIF has only two free parameters: The core-to-cladding refractive index contrast and the core radius. We calculate the respective refractive indexes using appropriate Sellmeier data in a range of wavelengths (850--1400 nm) for a specified Germanium doping concentration in the interval \mbox{0.0~\%--7.9~\%} \cite{Adams_1981}; the cladding is assumed to be pure silica; the core radius is varied in the interval 3.0 $\upmu$m--7.5 $\upmu$m. For every combination of doping concentration and core radius, we use Eq. \eref{eq_DB} to solve $\Delta \beta = 0$ for the signal frequency in the \LPo-mode using $\lambda_\p = 1064$ nm. The total pump power and its modal distribution are not important for phase matching because the nonlinear phase shift is small compared to the linear phase mismatch. \par
\begin{figure}
\centering\begin{subfigure}{0.49\linewidth}  \centering
\includegraphics[scale=1]{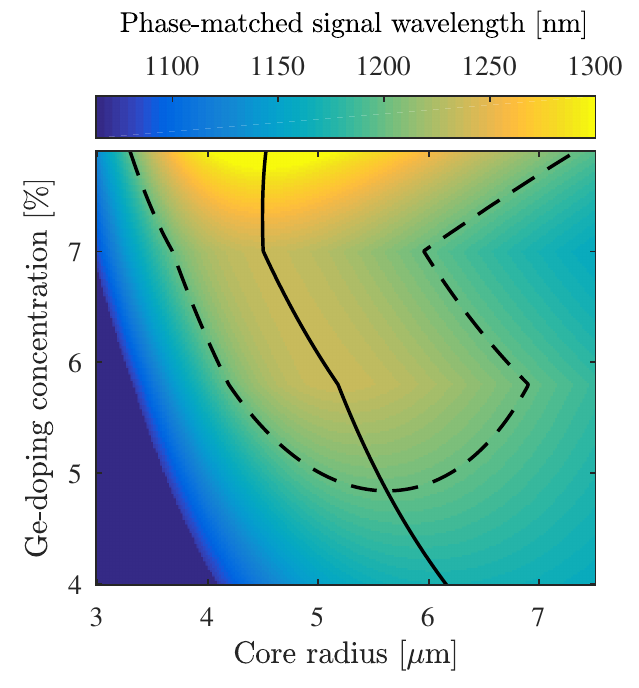}
\caption{}
\end{subfigure} 
\begin{subfigure}{0.49\linewidth}  \centering
\includegraphics[scale=1]{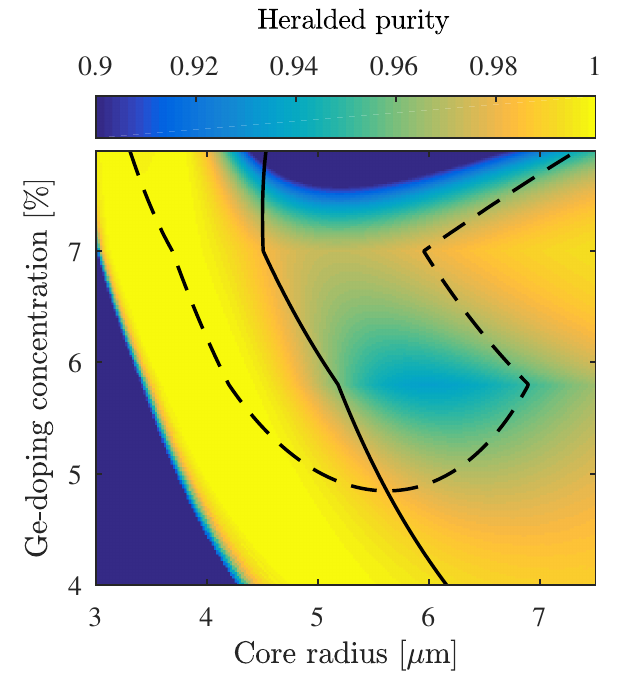}
\caption{}
\end{subfigure}
\caption{(a) Phase-matched signal wavelength and (b) heralded single-photon purity versus core radius and Ge-doping concentration of a SIF; the solid black line denotes all positions where the phase-matched signal wavelength has a maximum in the core radius; the dashed black line is a contour line that represents a separation of 32 THz between the pump at $\lambda_\p = 1064$ nm and the phase-matched wavelength.} \label{fig_2D_graphs}
\end{figure}
Figure \ref{fig_2D_graphs}(a) shows the phase-matched signal wavelength versus core doping concentration and fiber core radius; the dashed black line is a contour of the wavelength at which the signal is separated by more than $32$ THz from the pump; inside this contour, Raman scattering is negligible. An important feature of Fig. \ref{fig_2D_graphs}(a) is that for every doping concentration there is a core radius at which the signal wavelength has a maximum; this set of doping concentration and core radii is marked by the solid black line. Figure \ref{fig_2D_graphs}(b) shows the spectral purity of the heralded photons calculated from Eq. \eref{eq_Purity} corresponding to the signal wavelengths in Fig. \ref{fig_2D_graphs}(a); the Raman contour and the line of maximum signal wavelength are again indicated. The purity is seen to be anti-correlated with the signal wavelength (when the purity is high, the signal wavelength is small and vice versa). This property follows from the discussion above: large wavelength separations requires the parallelogram in Fig. \ref{fig_RIGV_sketch}(b) to be skewed, which inherently reduces the purity.\par
\begin{figure}
\centering\begin{subfigure}{0.49\linewidth}  \centering
\includegraphics[scale=1]{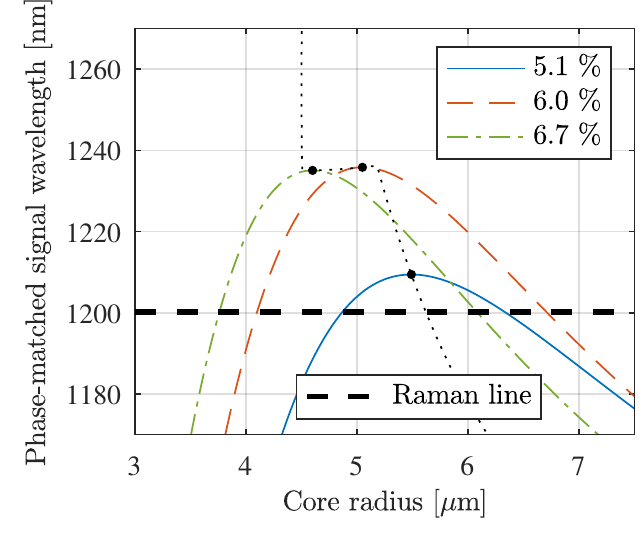}
\caption{}
\end{subfigure} 
\begin{subfigure}{0.49\linewidth}  \centering
\includegraphics[scale=1]{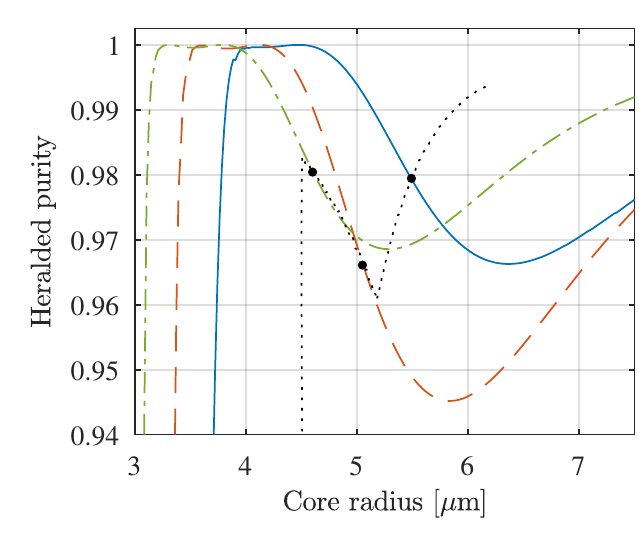}
\caption{}
\end{subfigure}
\caption{(a) Phase-matched signal wavelength and (b) heralded single-photon purity versus core radius for selected values of the doping concentration, i.e. horizontal cross sections of Fig. \ref{fig_2D_graphs}(a) and \ref{fig_2D_graphs}(b), respectively; the Raman line denotes the wavelength above which Raman scattering is negligible; in (a), the filled black dots mark the maximum on each curve; in (b), they denote the purity associated with each maximum in (a); the black dotted line in each plot represents the function values of the solid black lines in Fig. \ref{fig_2D_graphs}(a) and \ref{fig_2D_graphs}(b), respectively; the legend in (a) applies also to (b).} \label{fig_1D_graphs}
\end{figure}

Figures \ref{fig_1D_graphs}(a) and \ref{fig_1D_graphs}(b) show horizontal cross sections of Fig. \ref{fig_2D_graphs}(a) and \ref{fig_2D_graphs}(b), respectively, at selected doping concentrations, 5.1 \% (solid blue), 6.0 \% (dashed red), and 6.7 \% (dash-dotted green); the black dots in Fig. \ref{fig_1D_graphs}(a) mark the maximum on each curve and in Fig. \ref{fig_1D_graphs}(b) they mark the purity at this maximum signal wavelength, thus playing the same role as the solid black line in Fig. \ref{fig_2D_graphs}(a) and \ref{fig_2D_graphs}(b). The dotted black line in Fig. \ref{fig_1D_graphs}(a) and \ref{fig_1D_graphs}(b) shows the values that the maxima marked by the black dots follow when varying the doping concentration, i.e. the signal wavelengths and purity along the solid black line in Fig. \ref{fig_2D_graphs}(a) and \ref{fig_2D_graphs}(b), respectively.

The curve with doping concentration 5.1 \% has maximum slightly above the Raman contour but a high purity of 0.98 is observed. The curve with doping concentration 6.0~\% has maximum at a larger wavelength, further from the Raman contour, but the purity is less than 0.97. The curve with doping concentration 6.7~\% has approximately the same maximum wavelength as the 6.0~\% curve but the purity is as high as for the 5.1~\% curve. To avoid any residual spontaneous Raman scattering above 32 THz from the pump, it is advantageous to choose the doping concentration with phase matching at the furthest separation from the pump. Also, the loss in silica fibers reduces significantly with increasing wavelength in this region. We therefore choose a doping concentration of 6.7~\% to investigate further.

The importance of the maximum in the phase-matched signal wavelength is evident: at the maximum value, the signal wavelength and the purity become independent of core radius to first order. As we show below, this property enables photon-pair generation of high purity with realistic degrees of CRF in a silica fiber. \par

%
\begin{figure}
\centering\begin{subfigure}{0.325\linewidth}  \centering
\includegraphics[scale=0.97]{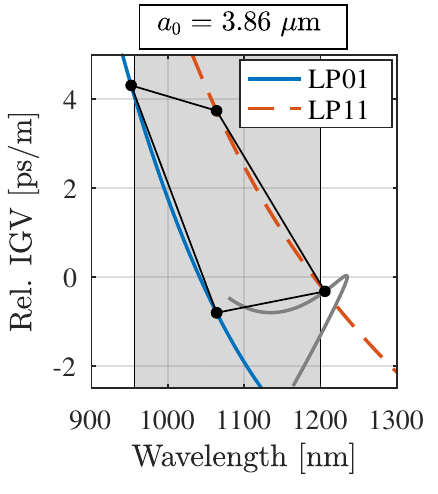}
\caption{}
\end{subfigure} 
\begin{subfigure}{0.325\linewidth}  \centering
\includegraphics[scale=0.97]{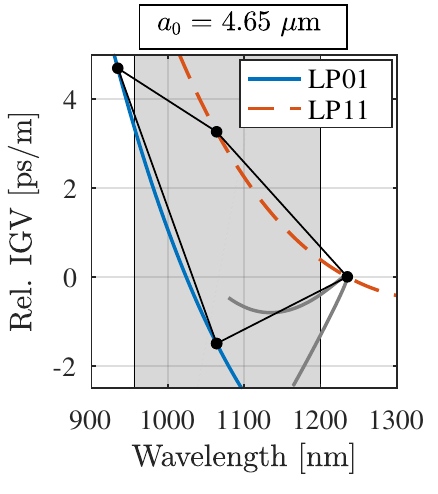}
\caption{}
\end{subfigure}
\begin{subfigure}{0.325\linewidth}  \centering
\includegraphics[scale=0.97]{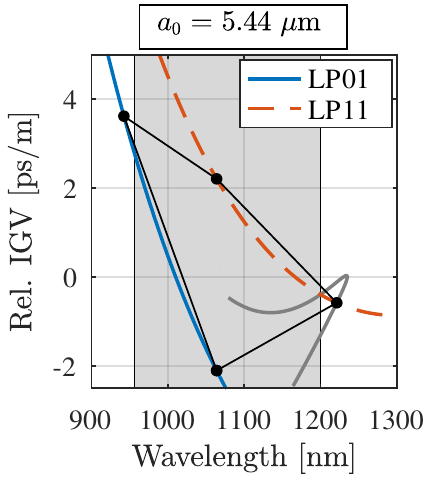}
\caption{}
\end{subfigure}
\caption{Relative IGV versus wavelength for the \LPo and \LPt-modes in a fiber with core radii (a) $a_0 = 3.86$ $\upmu$m, (b) $a_0 = 4.65$ $\upmu$m, and (c) $a_0 = 5.44$ $\upmu$m and doping concentration of 6.7 \%; the gray area marks the Raman active zone of 32 THz separation from the pump wavelength.} \label{fig_cartoon}
\end{figure}
The mechanism behind the formation of the maximum signal wavelength is illustrated in Fig. \ref{fig_cartoon}, where the IGV of the \LPo and \LPt-modes of a SIF with doping concentration 6.7 \% and three different core radii are shown; the skewed parallelograms show the sets of wavelengths for which phase matching is achieved at each core radius; the gray-filled area stretches 32 THz on either side of the pumps, thus marking the zone of Raman contamination. For increasing core radius from Fig. \ref{fig_cartoon}(a)$\rightarrow$\ref{fig_cartoon}(b), the IGV curve of the \LPt-mode increases relative to that of the \LPo-mode, which separates the signal/idler further from the pump. Further increasing the core radius from Fig. \ref{fig_cartoon}(b)$\rightarrow$\ref{fig_cartoon}(c) reduces the IGV separation of the \LPo{}  and \LPt-modes{}  again, and thus the signal wavelength moves closer to the pumps. The solid gray line marks the position of the signal wavelength in these plots when varying the core radius from \mbox{3.0 $\upmu$m} to \mbox{7.5 $\upmu$m}; the maximum wavelength on the gray line is clearly visible at $\lambda_{\rm PM} \approx 1235$ nm for a core radius of $a_0 = 4.65$ $\upmu$m. The normalized frequency of a fiber with this core radius and the chosen doping concentration at 1064 nm is 4.66, so the modes LP$_{21}$ and LP$_{02}$ are also guided; however, in fibers of less than 100 m, which we consider in this work, linear mode coupling is negligible. The cut-off wavelength of the \LPt-mode is approximately 2230 nm, which ensures that the mode is well guided and not sensitive to macro-bending losses.


\subsection{Heralded purity for varying core radii and pump-pulse durations}

Due to the scale independence of the joint-state factorability, there are only two parameters determining the impact of CRF on the heralded purity. The first is the ratio of fluctuations in phase-matched frequency to the spectral width of the joint state, which is investigated in this section. The second is the ratio of interaction length to the correlation length of fluctuations, which is considered in the next section. The first ratio is to second order
\begin{equation}
\label{omega_ratio}
\frac{\Delta \omega_\mathrm{PM}}{\Delta\omega_\mathrm{state}} = \frac{\ud \omega_\mathrm{PM}}{\ud a} \Big |_{a = a_0} \sigma_a T_\p + \frac{1}{2}\frac{\ud^2 \omega_\mathrm{PM}}{\ud a^2} \Big |_{a = a_0} \sigma_a^2 T_\p.
\end{equation}
If $\omega_\mathrm{PM}$ is not maximal with respect to $a$, the first term dominates and the purity depends mainly on the product $\sigma_a T_\p$. In the following we consider a 1-ps pulse corresponding to a fiber length of $L = 2.8~\mathrm{m}$ to ensure full walk-through, but the scaling of Eq. \eref{omega_ratio} makes the results more general.

\begin{figure}[ht]
\centering\begin{subfigure}{0.49\linewidth}  \centering
\includegraphics[scale=1]{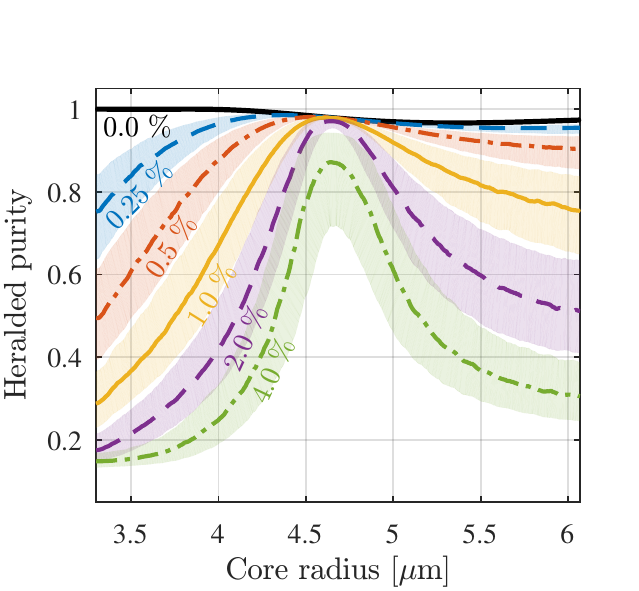}
\caption{}
\label{fig:Purity_vs_CoreRadius}
\end{subfigure} 
\begin{subfigure}{0.49\linewidth}  \centering
\includegraphics[scale=1]{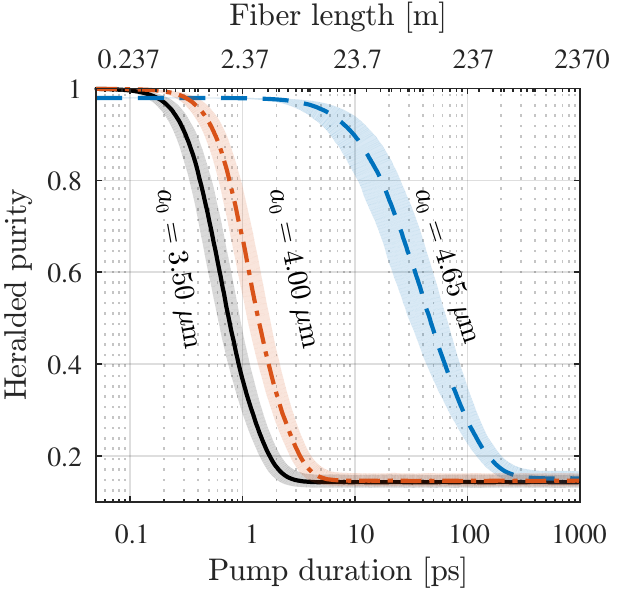}
\caption{}
\label{fig:Purity_vs_PulseDuration}
\end{subfigure}
\caption{Median heralded purity as a function of \textbf{(a)} core radius for different fluctuation strengths, with the value of $\sigma_a/a_0$ indicated next to each curve and 1-ps-pump pulses, and \textbf{(b)} pump pulse duration for different core radii for 1~\% CRF. Quantiles of the distribution are indicated such that half of the purities fall within the shaded region.} \label{fig:SpectralRatio}
\end{figure}

To quantify the impact of CRF on the heralded single-photon purity, we simulated for each of 5 different fluctuations strengths $10^6$ fibers with varying mean core radii, as described in Sec. \ref{sec:fluctuations}. For each fiber the purity was calculated and the results are shown in Fig. \ref{fig:SpectralRatio}(a). The running median and quantiles were taken over $50~\mathrm{nm}$ (corresponding to approximately 12000 points).
The figure clearly illustrates that the core radius $a_0 = 4.65~\upmu \mathrm{m}$, where the phase-matching condition is stable in wavelength, is significantly more robust to CRF. At this core radius the purity remains almost maximal even for unrealistically large fluctuations of $2~\%$, whereas for almost all other core radii the purity is significantly decreased for fluctuations as small as $0.25~\%$. The decrease is most significant at the core radius where the purity is largest in the absence of CRF, at which one might naively design a fiber. 

The impact of CRF on heralded purity as a function of pump-pulse duration and fiber length (to allow full pump walk-through) for a fluctuation strength of $\sigma_a/a_0 = 1~\%$ is shown in Fig. \ref{fig:SpectralRatio}(b). The simulation was performed for $10^6$ pulse durations and running quantiles were taken over $20000$ points. The core radius that gives a stable phase-matching condition is much more robust to CRF and gives high purities for pulses with durations up to $\sim 3~\mathrm{ps}$. In contrast, the two other radii displayed in the figure become unusable for pump pulses as short as $200~\mathrm{fs}$ and $100~\mathrm{fs}$, respectively. As we show later, the reduction in purity for increasing pump duration and fiber length is mainly caused by the smaller spectral extent of generated photons resulting in higher sensitivity to changes in the phase-matching condition, and not because the fiber is longer relative to the correlation length of fluctuations.

\subsection{Impact of correlation length on heralded purity and intersource visibility}
\begin{figure}[!t]
\centering\begin{subfigure}{0.49\linewidth}  \centering
\includegraphics[scale=1]{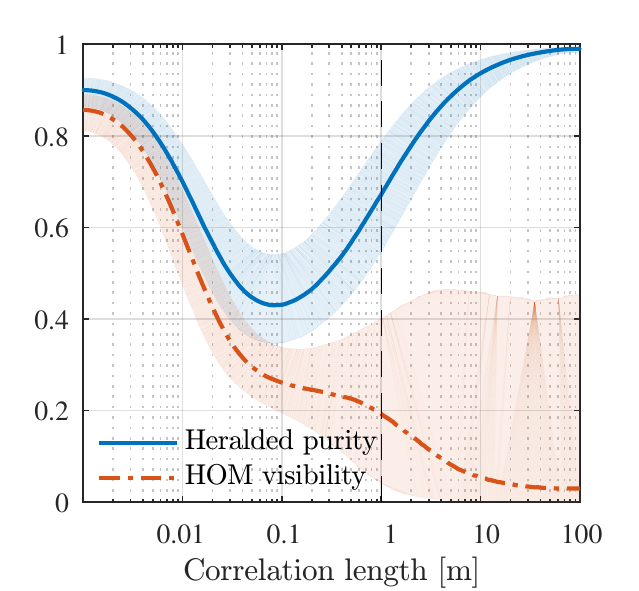}
\caption{}
\end{subfigure} 
\begin{subfigure}{0.49\linewidth}  \centering
\includegraphics[scale=1]{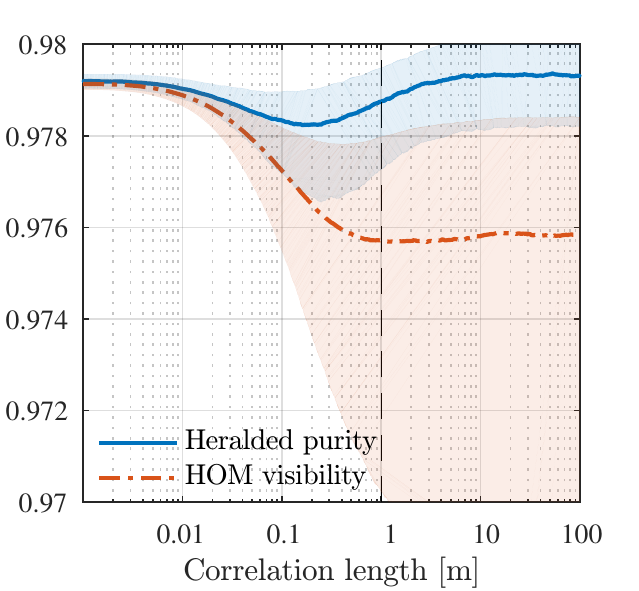}
\caption{}
\end{subfigure}
\caption{\textbf{(a)} Median purity of heralded photons from a source and HOM visibility of two sources with $a_0 = 4.0~\upmu\mathrm{m}$ as a function of correlation length. \textbf{(b)} Same as \textbf{(a)}, but with $a_0 = 4.65~\upmu\mathrm{m}$ where the phase matching is stable. Half of the simulation results fall within the shaded regions and the dashed line indicates the correlation length used in previous simulations.} \label{fig:correlation_length}
\end{figure}

Although the correlation length of fluctuations is in most cases not a controllable parameter, the source performance nevertheless depends on this parameter. The only length scales in this problem are the pump-pump collision length, $l_\mathrm{coll} = T_\p/|\beta_{1\p} - \beta_{1\q}|$, and the correlation length, $l_\mathrm{corr}$, of fluctuations, and thus the dependence on correlation length is only controlled by the ratio
\begin{equation}
\frac{l_\mathrm{coll}}{l_\mathrm{corr}} =  \frac{T_\p}{l_\mathrm{corr} |\beta_{1\p} - \beta_{1\q}|}.
\end{equation}
We again consider a 1-ps pump pulse, but the dependence on $l_\mathrm{corr}$ is similar for other pulse durations by a scaling with $T_\p$. We characterize the impact by considering the heralded single-photon purity $P = \Tr(\hat{\rho}_\s^2)$ and the intersource two-photon interference visibility $V = \Tr(\hat{\rho}^{(1)}_{\s}\hat{\rho}^{(2)}_{\s})$ as given by Eq. \eref{eq:visibility}. Figure \ref{fig:correlation_length} shows the purity and visibility as a function of the correlation length at $a_0 = 4.00~\upmu\mathrm{m}$ in Fig. \ref{fig:correlation_length}(a) and at the fluctuation-resistant core radius $a_0 = 4.65~\upmu\mathrm{m}$, both with CRF of 1~\%. The results are based on simulations of $10^6$ fibers with the quantiles taken over $10^5$ points due to large variation in the observed visibilities.
Comparing Figs. \ref{fig:correlation_length}(a) and \ref{fig:correlation_length}(b), we observe again a significant difference in resistance to fluctuations (notice the purity axes are different by a factor of 100). The purity is severely reduced for 1~\% fluctuations at $a_0 = 4.00~\upmu\mathrm{m}$ while an insignificant reduction is observed for the stable core radius. However, while the severity of the reduction differs greatly between the two radii, the variation has a similar character despite the very different value of the other important ratio $\omega_\mathrm{PM}/\omega_\mathrm{state}$. \par
When the correlation length is small compared to the interaction length ($l_\mathrm{corr} \ll |\beta_{1\p}-\beta_{1\q}|/T_\p$), the variation becomes completely uncorrelated along the fiber. While this kind of variation has an impact on the state and reduces the purity, depending on the fluctuation strength $\sigma_a$, it does not displace the whole spectral distribution in frequency, resulting in a visibility very near the purity limit, which is illustrated in the figure for both core radii. For large correlation lengths ($l_\mathrm{corr} \gg |\beta_{1\p}-\beta_{1\q}|/T_\p$), the core radius fluctuates very little throughout the fiber, resulting in photons of higher purity than for fast fluctuations, since the JSA is not distorted. However, fiber-to-fiber variations of this constant value, limits the achievable interference visibility between two different sources, due to photon distinguishability arising from a spectral displacement of the whole two-photon state. In practical applications this can be avoided by selecting identically performing sources or by using pieces of fiber cut from the same longer fiber. Interestingly, in terms of purity there is a worst correlation length $l_\mathrm{coll}/l_\mathrm{corr} = 2.55$ ($L/l_\mathrm{corr} = 28.8$) for $a_0 = 4.00~\upmu\mathrm{m}$ and $l_\mathrm{coll}/l_\mathrm{corr} = 1.27$ ($L/l_\mathrm{corr} = 14.4$) for $a_0 = 4.65~\upmu\mathrm{m}$. In this intermediate case, the fluctuations are rapid enough to change the value over the interaction length, but slow enough to not average out in the sense seen in the limit $l_\mathrm{corr} \to 0$. 

\FloatBarrier

\subsection{Discussion}

Our analysis shows that fabrication fluctuations directly impose a limit on the obtainable purity, with the main determining parameter being the ratio of the joint state spectral width to the mean fluctuations in phase-matching frequency. In the multi-mode scheme we propose in this work where the phase-matched signal wavelength is stable at a specific core radius, this imposes an upper limit for the pump-pulse durations whereas in the asymmetrically group-velocity matched scheme used in PCFs~\cite{halder2009nonclassical,cohen2009tailored,soller2010bridging}, a limit is imposed on the maximal length of fiber that can be used. In the latter scheme, this directly limits the obtainable purity since it approaches unity asymptotically as the fiber length increases. It has been proposed to splice together suitable pieces of PCF to create a longer homogeneous fiber~\cite{cui2012spectral}, but this may be impractical due to the short correlation length of geometric fluctuations~\cite{francis2016characterisation}. As was the case here, it may be possible to design other waveguide platforms, such as PCFs, to have a stable phase-matching condition with respect to the geometric imperfections although such an analysis is clearly much more complex than for a simple SIF. \par
In this work we neglected the impact of nonlinear phase modulation (NPM) which may impress a practical limitation on the single-photon purity~\cite{bell2015effects}. The scheme considered in this work has been shown to be fairly resistant to NPM~\cite{christensen2016temporally}, but the higher degree of asymmetry in group velocities and the different nonlinear overlaps on the multi-mode platform makes it unclear if the effects are significant. A detailed study of this would be desirable in future work. It should be noted, however, that the impact of nonlinear phase modulation is in principle avoidable by simply compromising on pair-generation rate, whereas the impact due to fabrication imperfections is present at all generation rates. \par
Finally, we note that there are unexploited tunable parameters in this scheme such as pump-pulse durations and power distributions, which have been ignored to highlight the impact of dispersion fluctuations. Before an experimental realization, further optimization should be performed to achieve an acceptable compromise between operating wavelengths (ideally close to a signal wavelength of 1550 nm), heralded purity and resistance to fabrication imperfections and nonlinear phase modulation. It is likely that a more advanced fiber design than the simple SIF considered here is necessary to achieve good performance on all parameters.

\section{Conclusion}

We developed a model for describing the two-photon state generated by spontaneous four-wave mixing under stochastic variation of waveguide parameters. This model was used to characterize the impact of core-radius fluctuations in a step-index fiber relying on mode-induced walk-off to produce pure heralded single photons. The fiber source was designed to generate pure single photons with a large spectral separation from the pump fields to avoid contamination from spontaneous Raman scattering and facilitating filtering of the strong pump fields. \par 
We have demonstrated how the impact of core-radius fluctuations is determined only by two ratios: the ratio of fluctuations in the phase-matched frequencies to the spectral width of the two-photon states and the ratio of interaction length to correlation length of fluctuations. The latter was shown to affect the purity as well as the intersource two-photon interference visibility. It was found that the impact was the most severe when the correlation length is comparable to the interaction length. The ratio of phase-matching fluctuations to spectral width of the two-photon state was found to be the most significant. Even for fast 1-ps pump pulses a severe reduction in purity was observed for fluctuations in core radius below $0.5~\%$. However, it was discovered that a core radius that gives a core-radius-independent phase-matching condition to first order exists. At this stable core radius, a very significant improvement in robustness to fluctuations was observed. This allows a 1~ps pump pulse to generate very pure photons ($>98\ \%$) even for unrealistically large fluctuations (above 1~\% standard deviation). The fiber-based higher-order-mode scheme presented in this work hence lays the theoretical foundation for generation of pure single photons by intermodal four-wave mixing in a simple step-index fiber. \par
Although we here considered a specific scheme for pure single-photon generation, fabrication imperfections are present in all platforms and are likely to affect the spectrally tailored two-photon states in many other systems. We have shown how to guide the design of such a system to be resistant to fabrication imperfections, although the analysis may be much more complicated in more complex systems, where several parameters can vary independently.

\section*{Funding}
This work was supported by the Danish Council for Independent Research (DFF) (4184-00433).


\small
\bibliographystyle{ieeetr}

\end{document}